\documentclass[aps, prl, reprint, superscriptaddress, showpacs, floatfix]{revtex4-1}

\usepackage{graphicx}
\usepackage{amsmath}
\usepackage{epstopdf}

\begin{document}

\title{Diversity enabling equilibration: disorder and the ground state in artificial spin ice}

\author{Zoe Budrikis}
\email{zoe.budrikis@gmail.com}
\affiliation{School of Physics, The University of Western Australia, 35 Stirling Hwy Crawley WA 6009 Australia}
\affiliation{Istituto dei Sistemi Complessi, Consiglio Nazionale delle Ricerche, Via Madonna del Piano 10, 50019 Sesto Fiorentino, Italy}
\author{Paolo Politi}
\affiliation{Istituto dei Sistemi Complessi, Consiglio Nazionale delle Ricerche, Via Madonna del Piano 10, 50019 Sesto Fiorentino, Italy}
\affiliation{INFN Sezione di Firenze, via G. Sansone 1, 50019 Sesto Fiorentino, Italy}
\author{R. L. Stamps}
\affiliation{School of Physics, The University of Western Australia, 35 Stirling Hwy Crawley WA 6009 Australia}
\affiliation{SUPA, School of Physics and Astronomy, University of Glasgow, Glasgow G12 8QQ, United Kingdom}

\begin{abstract}
We report a novel approach to the question of whether and how the ground state can be achieved in square artificial spin ices where frustration is incomplete. 
We identify two sources of randomness that affect the approach to ground state: quenched disorder in the island response to fields and randomness in the sequence of driving fields. 
Numerical simulations show that quenched disorder can lead to final states with lower energy, and randomness in the sequence of driving fields always lowers the final energy attained by the system. 
We use a network picture to understand these two effects: disorder in island responses creates new dynamical pathways, and a random sequence of driving fields allows more pathways to be followed.
\end{abstract}

\pacs{75.50.Lk, 75.75.-c, 75.78.-n}

\maketitle

\begin{figure}
\centering
 \includegraphics[width=0.8\columnwidth]{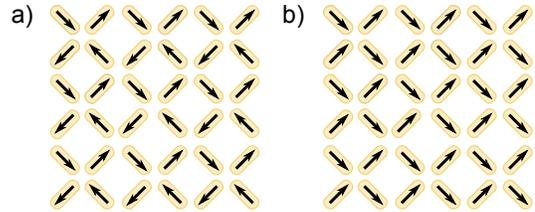}
  \caption{\label{array_geometry}
The array geometry studied, shown for $6\times6$ islands.
  (a) The ground state configuration in which spins are arranged in a microvortex configuration. The ground state is two-fold degenerate, with a global spin flip giving the other ground state configuration.
  (b) A saturated configuration, in which all spins have a positive projection onto an axis. There are four saturated configurations.
}
\end{figure}

The ability to control artificial spin ice~\cite{Wang:2006, Moller:2006, Nisoli:2007, Ke:2008, Remhof:2008, Budrikis:2010, Morgan:2010, Phatak:2011, Kohli:2011, Tanaka:2006, Qi:2008, Mengotti:2010, Ladak:2010, Mellado:2010, Li:2010} and related systems~\cite{Libal:2006, Libal:2009} has made them a useful testing ground for many types of physics. For example, the effective temperature formalism used for granular materials can be studied in the context of artificial spin ice~\cite{Mehta:1989, Nisoli:2007, Nisoli:2010}.
The submicron Ising-like magnetic islands of artificial spin ice are arranged so the dipolar inter-island interactions are frustrated, leading to a complex energy landscape with many states nearly degenerate.
In square artificial spin ices, the interactions within each vertex of the lattice of islands are inequivalent, leading to a well-defined, two-fold degenerate ground state, as shown in Fig.~\ref{array_geometry}(a). Large domains of ground state ordering have been observed in samples that have undergone thermal annealing during growth~\cite{Morgan:2010}. However, once grown, the islands are large enough to be athermal and dynamics are driven entirely by external fields.

One difference between athermal driven and thermal dynamics is that an external driving field acts uniformly on all islands, whereas thermally driven individual islands may reverse stochastically. A consequence of having global and deterministic driving -- rather than localized and random -- is that dynamics are constrained and even though the inequivalence of interactions lifts degeneracies in the energy landscape, the field driven dynamics cannot neccessarily evolve the ice to a low energy state. Experimentally, a sequence of fields applied to a saturated configuration (shown in Fig.~\ref{array_geometry}(b)) gives a final state with only short range ground state ordering~\cite{Wang:2006, Wang:2007, Ke:2008, Phatak:2011}.

Although the driving field is global, a spin's behavior is also determined by local factors: its interactions with other spins and its (intrinsic) switching field. Disorder in these affects the response to driving. 
Disorder has previously been shown experimentally to be important for the response of square artificial spin ice to fields~\cite{Kohli:2011} but until now, the mechanisms have been unclear.
Here, we focus on the simplest type of disorder, which is to allow the switching fields to vary in magnitude from island to island, as would result from roughness in the islands. This variation in switching fields does not modify the interactions, and frustration and degeneracies are unaffected. Instead, it modifies the way the system explores its energy landscape.
We study the effects of disorder using numerical simulations which give full control over island properties, and by mapping transitions onto a directed network which can be studied using network theoretic tools~\cite{Albert:2002, Newman:2003}. This latter approach provides a conceptual framework which can be applied to other systems.

We first describe the effect of switching field disorder on the final energy of the system, when a uniformly rotating protocol is used. 
Then, we interpret the results with the support of the network description, which is also the natural framework to study changes in the field protocol. 
Finally, we study field protocols with a random sequence of field angles, and we show they allow the system to achieve lower-energy final states than can occur under rotating protocols, in a way that is more robust against quenched disorder.

In our simulations, the magnetic islands are treated as dimensionless Ising spins $\vec S_i$ ($S=1$) that interact as point dipoles.
The energy of spin $i$ is given by its Zeeman energy, $-\vec{h}\cdot\vec{S}_i$, and the sum of dipolar interactions with all other spins:
\begin{equation}
\begin{split}
E_{\mathrm{dip}}^{(i)} &= -\vec h_{\mathrm{dip}}^{(i)}\cdot \vec{S}_i \\
	&= \frac{1}{4 \pi \mu_0} \sum_{j\ne i} \biggl(
	\frac{\vec{S}_i \cdot \vec{S}_j}{r_{ij}^3}
	- 3 \frac{(\vec{S}_i \cdot \vec{r_{ij}}) (\vec{S}_i \cdot \vec{r_{ij}})}{r_{ij}^5}
	\biggr).
\end{split}
\end{equation}
As in previous works \cite{Moller:2006, Budrikis:2010}, spin $i$ can flip if
\begin{equation}
\label{switching}
\vec h_{\mathrm{tot}}^{(i)}\cdot \vec S_i < -h_c^{(i)},
\end{equation}
where $h_{\mathrm{tot}}^{(i)} = \vec{h} + \vec h_{\mathrm{dip}}^{(i)}$ and $h_c^{(i)}$ is the island's intrinsic switching field. 

Evolution under an external field $\vec{h}$ proceeds by randomly selecting a spin satisfying criterion \eqref{switching}, flipping it, then checking the switching condition again for all spins, until no further flips are possible.
We denote the transition under a field $\vec{h}$ from initial state $i$ to final state $f$ by $i \to f$. (More than one final state may be possible; alternatively $f$ may be the same state as $i$ if no spins can flip.)
The transition $i\to f$ is the building block of the simulated dynamics and the network picture, discussed below.

In the absence of disorder, all islands have the same switching field. 
We implement disorder by drawing the $h_c^{(i)}$ from a Gaussian distribution characterized by standard deviation $\sigma$.
We work in units where $1/(4\pi\mu_0)$ and the nearest-neighbor distance are set to unity, and the mean $h_c$ value is $11.25$.

\begin{figure}
\centering
 \includegraphics[width=0.85\columnwidth]{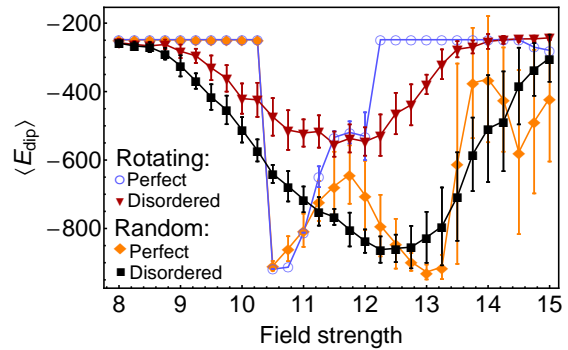}
  \caption{\label{energy_rotating_random}
Disorder in island switching fields or the sequence of applied fields allows the system to reach final states with low energy. 
As indicated by the legend, symbols represent rotating and random field protocols acting on perfect and disordered systems.
  Averages are made over 20 disorder realizations. Error bars represent one standard deviation.
}
\end{figure}

In a perfect system with edge geometry as shown in Fig.~\ref{array_geometry}, a rotating field with constant amplitude can induce non-trivial dynamics only in a narrow range of field amplitudes, $\Delta h \simeq 2$. This is shown for a $20\times20$ spin system as blue circles in Fig.~\ref{energy_rotating_random}. As discussed in Ref.~\cite{Budrikis:2010}, smaller fields are
unable to flip spins and larger fields force the magnetization to track the field. Dynamics always start at array edges where spins have fewer neighbors.
The minimum of $E_{\mathrm{dip}}(h)$ at $h\simeq 10.5$ is due to a very regular and specific process of spin flips that ``invade'' from the array edges. 

Let us now consider the effect of switching field disorder, with $\sigma=1.875$. This disorder strength is consistent with experimental data for the square lattices studied here~\cite{Morgan:2011} and disconnected kagome ices~\cite{Mengotti:2010, Ladak:2010}.
Disorder allows dynamics to start inside the array at sites where ``loose'' spins with smaller switching fields are located. Similar disorder effects have been seen in simulations of vortices in nanopatterned superconductors~\cite{Libal:2009}.
The uniformly rotating field
protocol now produces the curve $E_{\mathrm{dip}}(h)$ given by inverted triangles
in Fig.~\ref{energy_rotating_random}.
The increase in $\Delta h$ indicates that the available configuration space is enlarged: dynamical processes forbidden in the perfect system -- such as the flipping of isolated spins in the array bulk -- are allowed when disorder is present.
However, disorder can also destroy transitions. For example, the orderly process of spin flips at $h\simeq 10.5$ in the perfect system cannot occur in the disordered system. This leads to higher energy at these fields.

Up to now, the discussion has focussed on the dynamics of spins in the array. However, an alternative viewpoint -- alluded to in the discussion of transitions $i \to f$ -- is to see the action of the applied field as ``transporting'' the system from one spin configuration state to another, in a path through the space of all configurations. This picture can be interpreted as a network, in which network nodes are spin configurations and a directed link $(i \to f)$ exists if 
applying some field to configuration $i$ allows it to transition to configuration $f$, according to the dynamics described above.
The advantage of this approach is that it allows us to discuss dynamics in a way that is not restricted to a particular field protocol.
Related approaches to mapping dynamics onto networks have proved fruitful in the study of other geometrically frustrated systems~\cite{Han:2009, Han:2010} and the random field Ising model~\cite{Bertotti:2007, Bortolotti:2008, Bortolotti:2010}.

We restrict our considerations to fields of strength $h=11.5$. Using a different field amplitude would give a different network.
In principle, the field can take any angle in $(0, 2\pi]$ but we consider only the discrete set of field angles $\theta = n \pi/128$, $n=1,2,\ldots256$. As seen in the Supplementary Material~\cite{SM1}, this set of field angles can be expected to give results close to the limit of continuous $\theta$. For each configuration (node), transitions are calculated at all field angles $\theta$.
A field protocol $\{\theta_1, \theta_2, \ldots\}$ corresponds to a path on the network where at each step of the path only a link corresponding to the angle $\theta_i$ can be followed.

Since the number of configuration states increases exponentially with the size of the lattice, we consider a small, feasible to analyze, $4\times4$ array, 
with 16 islands and $2^{16}$ total configurations. We are able to obtain the exact network representation of the dynamics of such a system, via enumeration.
Although the dynamics of the $4\times4$ system are not exactly the same as those of the $20\times20$ system, they are governed by the same basic rules, and results for the smaller system are applicable to the larger one~\cite{SM1}.
The networks are stored as adjacency matrices.
$A_{if}=1$ if the transition $i \to f$ is allowed for some $\theta$, $A_{if}=0$ otherwise. 
These are sparse matrices, with $\sim10^6$ non-zero entries in a $2^{16}\times2^{16}$ matrix.
Disorder is implemented in a similar way to the simulations. However, because of the small number of islands, the actual mean of the $h_c$ values drawn from the distribution is $11.25+\Delta$, where $\Delta$ can be large enough to affect the network properties. We subtract $\Delta$ from each $h_c$ value to fix the mean at $11.25$. Each disorder realization gives a different network, but we find similar network properties for disorder realizations with similar standard deviations in $h_c$.

\begin{figure}
\centering
 \includegraphics[width=0.8\columnwidth]{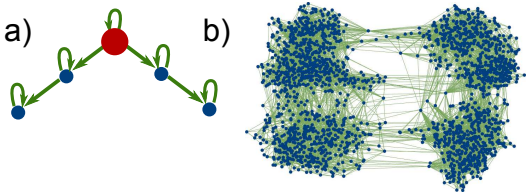}
  \caption{\label{subnetworks} 
  The subnetwork of states that can be accessed from the $+x$ saturated configuration for (a) a perfect system (where the saturated configuration is the large red node) and (b) a typical disordered system.
}
\end{figure}

Fig.~\ref{subnetworks}(a) shows the subnetwork of states reachable from the $+x$ saturated configuration in a perfect system. All the links that exist for $h=11.5$, $\theta=n\pi/128$ are shown. 
The most striking feature of the subnetwork is that it is very limited, containing only five nodes. 
The saturated configuration (indicated by the large red node) can evolve into one of two states, each of which can evolve into a single state, giving two possible final states for the system. Self-loops indicate that certain field angles are unable to induce transitions.
There are no ``return'' paths -- transitions are irreversible.
Fig.~\ref{subnetworks}(b) shows the subnetwork of states reachable from the same node, for a typical disorder realization. The subnetwork contains 1814 nodes, three orders of magnitude more than the perfect system. In this subnetwork, there are paths into the saturated configuration, as well as out of it. Other realizations of disorder give similar results, with the number of configurations reachable from the saturated state ranging from 1758 to 12516 in a study of ten disorder realizations.

\begin{figure}
\centering
 \includegraphics[width=0.75\columnwidth]{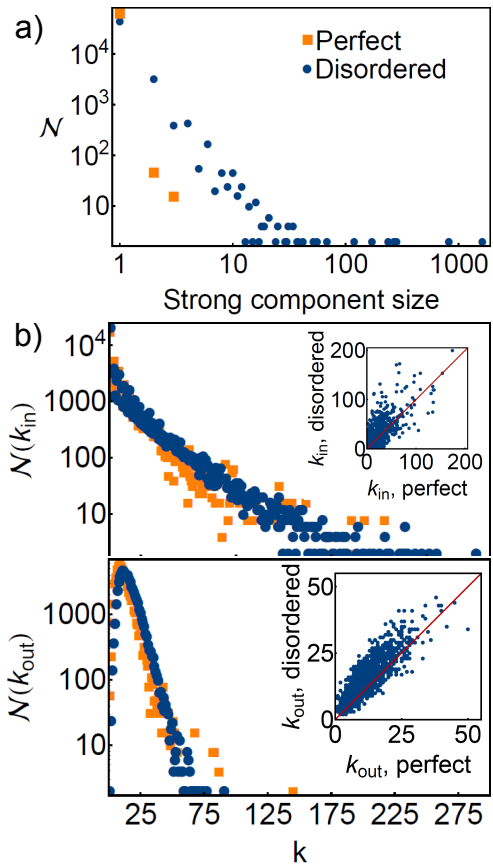}
  \caption{\label{networks} 
  (a) The distributions of strongly connected component sizes for the perfect (orange/light gray squares) and disordered (blue/dark gray circles) systems. 
  (b) The in- and out-degree distributions for the perfect (orange/light gray squares) and disordered (blue/dark gray circles) systems.
  Insets: The in- and out-degree of 1000 randomly chosen configurations, in the disordered \textit{vs} the perfect system. The red line separates those nodes whose degree is increased from those whose degree is decreased by disorder.}
\end{figure}

Turning to the network as a whole, we see that disorder ``re-wires'' the network significantly, increasing the total number of links from 736,720 to 949,216. This is not simply the addition of links: disorder removes 358,934 links from the perfect system, and creates 571,430 links.
As already mentioned, one result of this re-wiring is an increase in the reversibility of dynamical pathways. This can be described quantitatively using strongly connected components (SCCs). In a directed network, if nodes $A$ and $B$ are in the same SCC, there exists a path from $A$ to $B$ and vice-versa. The SCCs are determined using the algorithm in Ref.~\cite{Tarjan:1972}, as implemented by the software package $Mathematica$. Fig.~\ref{networks}(a) shows the distributions of SCC sizes for perfect and disordered systems. In the perfect system, the largest SCC contains 3 nodes and the saturated configurations are in SCCs of size 1. 
 In the disordered system the largest SCC contains 1628 nodes, and includes the saturated configurations. In terms of dynamics, this means that in the disordered system there exists a sequence of fields to take a saturated configuration to any one of 1627 configurations, and also a sequence back to saturation. Other disorder realizations give maximum SCC sizes ranging from 606 to 12386.

On the other hand, as seen in Fig.~\ref{networks}(b), the in- and out-degree distributions are not affected much by disorder. 
In both perfect and disordered systems, most nodes have a low in-degree, indicating that most configuration states can only be reached from a small number of other states. The peak in the out-degree distribution is at $k_{\mathrm{out}}\sim10$, with only a few nodes having very low out-degree (stable configurations) or very high out degree (unstable configurations). As seen in the insets to Fig.~\ref{networks}(b), nodes that have low (high) degree in the perfect system typically also have low (high) degree in the disordered system. The degree is increased more often than it is decreased because disorder adds links to the system which are distributed among the nodes.

While diversity in switching fields can open new links between states, a link
$i\to f$ is active only for certain orientations of the field,
$\theta_{i\to f}^{\min} <\theta< \theta_{i\to f}^{\max}$.
If the system is in state $i$, but the applied field is not at an angle
in the active interval, the link is not ``operative''. In this way, the field protocol is essential in determining dynamics. Changing the protocol can dramatically modify the final configurations attained.

\begin{figure}
\centering
 \includegraphics[width=0.85\columnwidth]{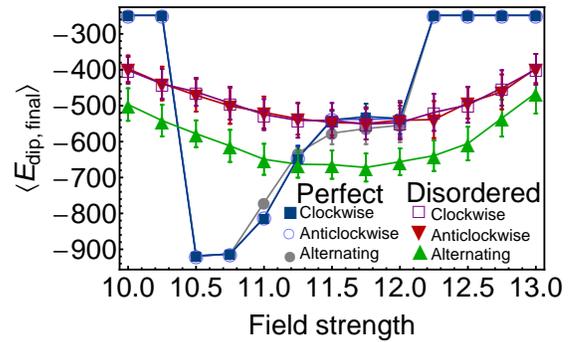}
  \caption{\label{reversing_not} 
  For an undisordered system, clockwise rotating fields, anticlockwise rotating fields, and a field protocol that alternates sense of rotation every 2 cycles all give the same final energy.
  When disorder is included,
  the clockwise and anticlockwise rotating fields 
  give the same results on average, but the alternating field protocol 
  leads to a lower final energy. Symbols for each protocol and system are indicated in the legend. Averages are made over 100 disorder realizations and error bars represent one standard deviation.}
\end{figure}

We now show evidence of the importance of field protocol by simulating alternative protocols. The first example, examined in Fig.~\ref{reversing_not}, takes advantage of the way disorder breaks symmetry between clockwise (CW) and anticlockwise (ACW)
rotations.
In the perfect system, a uniform rotation protocol gives the
same result whether the rotation is CW or ACW, because of reflection symmetry.
When disorder is introduced, the two senses of rotation
are no longer equivalent for any particular realization of disorder, but symmetry is restored when we average over disorder, as seen in Fig.~\ref{reversing_not}. 
We can go a step further and consider a
protocol that samples both senses of rotation by
repeatedly making a number of rotations CW and then ACW.
In the perfect system, this alternating CW/ACW protocol is equivalent to one with a single sense of rotation. In a disordered system, it leads to lower energy states with greater ground-state ordering.
The key point is that this simple modification of the uniform protocol, and the presence of disorder,
allow the system to avoid trapping by higher-energy local minima.

The alternating CW/ACW protocol breaks the regular sequence of applied fields to allow the system to traverse regions of its phase space network inaccessible to the uniform rotating protocol. 
Even more effective is a random sequence of field orientations, corresponding to a random walk on the network. The randomness introduced by this protocol is extrinsic and controllable, unlike the randomness introduced by switching field disorder. Results for random protocols are shown in Fig.~\ref{energy_rotating_random}. 
The diamonds show results for a perfect ice, and the squares show results for a disordered ice. In both cases the final energy is lower than that reached by a uniformly rotating protocol in the same system, and the fraction of vertices in a ground-state configuration is higher (not shown). The large error bars for $h\gtrsim11.5$ indicate the spread in final configurations attained, corresponding to a variety of pathways taken on the underlying dynamics network. Indeed, for $h\gtrsim13.5$, the simulations do not attain a single final state, and the averages of Fig.~\ref{energy_rotating_random} are made over configurations attained after the arbitrary time of 2000 field applications.

It is interesting to contrast these results for an athermal ice to those presented recently for thermal ices. In Ref.~\cite{Morgan:2010}, experimentally observed large scale ordering 
was attributed to thermally driven processes occurring during growth stages where the islands were 
small enough to undergo thermal fluctuations.
Thermal fluctuations are often thought of as stochastic local fields. In the athermal system considered here, 
they are replaced by either a random sequence of global fields or frozen randomness in the form of a distribution of switching fields.
The unifying principle for these two types of randomness is the concept of links between configurations. The links followed depend on both the switching fields and the driving field. Diversity in either of these enables 
a wider exploration of configuration space.

This ability to achieve long range ground state ordering in the athermal system suggests a process similar in certain respects to the ``order by disorder'' proposed by Villain
for equilibrium systems~\cite{Villain:1980}.
In order by disorder phenomena, disorder lifts degeneracies in frustrated systems and a ground state, inaccessible in a pure system at zero temperature can be acheieved by introducing either quenched or thermal disorder.
This is an equilibrium phenomenon where fluctuations 
select a specific ground state
from an ensemble of degenerate states.
By way of contrast, our athermal nonequilibrium system without disorder has a single ground state
and strongly constrained dynamics.
While disorder does not make the ground state accessible \cite{Morgan:2011}, it does remove
constraints by adding links between states.
With the correct field protocol, these links can be followed, allowing the system to relax into a state with a high level of ground state ordering.

\begin{acknowledgments}
We thank D.~ben-Avraham, C.~Castellano, C.~H.~Marrows, J.~P.~Morgan and J.~Villain for helpful discussions, and O. Hovorka for pointing out Refs~\cite{Bertotti:2007, Bortolotti:2008, Bortolotti:2010}. P.P. thanks Gian Piero Puccioni for help in the visualization of spin dynamics. Z.B. and R.L.S. acknowledge the Australian Research Council and the Worldwide University Network for funding. Z.B. acknowledges funding from INFN and the Hackett Foundation. 
\end{acknowledgments}

\end{document}